\documentclass[a4paper,11pt]{article}
\usepackage[dvips]{graphics}
\usepackage[british]{babel}
\usepackage{amssymb,amsmath}
\usepackage[ansinew]{inputenc}
\usepackage{epsfig}
\begin{document}
\begin{center}
\Large\textbf{{How to spell out the epistemic conception of quantum states\footnote{I am grateful to Patrick Pl\"otz for very helpful comments on an earlier version of this paper and to Chris Timpson for two nice discussions at Oxford. Furthermore, I would like to thank two anonymous referees of \textit{Studies in History and Philosophy of Modern Physics} for useful remarks.}}}\vspace{0.5cm}\\
\normalsize Simon Friederich\vspace{0.5cm}\\
\texttt{friederich@uni-wuppertal.de}\vspace{0.2cm}\\
Universit\"at Wuppertal, Fachbereich C -- Mathematik und Naturwissenschaften, Gau\ss str. 20, D-42119 Wuppertal, Germany\footnote{Work carried out at the Institute of Theoretical Physics, University of Heidelberg}
\end{center}

\small{
\noindent The paper investigates the epistemic conception of quantum states---the view that quantum states are not descriptions of quantum systems but rather reflect the assigning agents' epistemic relations to the systems. This idea, which can be found already in the works of Copenhagen adherents Heisenberg and Peierls, has received increasing attention in recent years because it promises an understanding of quantum theory in which neither the measurement problem nor a conflict between quantum non-locality and relativity theory arises. Here it is argued that the main challenge for proponents of this idea is to make sense of the notion of a state assignment being performed \textit{correctly} without thereby acknowledging the notion of a \textit{true} state of a quantum system---a state it \textit{is in}. An account based on the epistemic conception of states is proposed that fulfills this requirement by interpreting the rules governing state assignment as \textit{constitutive rules} in the sense of John Searle.
}\vspace{0.3cm}\\

\vspace{0.3cm}\\

\section{Introduction}

The epistemic conception of quantum states is the view that quantum states do not describe the properties of quantum systems but rather reflect the state-assigning agents' epistemic relations to the systems. Although this idea is not at all new---it has its roots in the works of Copenhagen adherents Heisenberg and Peierls\footnote{See, for instance, \cite{Heisenberg}, Chapter 3, and \cite{Peierls}.}---, the most important attempts of refining it or working it out in detail have been made only in more recent years.\footnote{\label{fussnote}For studies defending a version of the epistemic conception of states and views in a similar spirit see \cite{FuchsPeres}, \cite{Merminnew}, \cite{Caves2002a}, \cite{Caves2002b}, \cite{Fuchs}, \cite{Pitowsky}, \cite{Bub}, \cite{Caves2007}, \cite{BubPit}, \cite{Spekkens}, \cite{FuchsSchack1}, \cite{FuchsSchack2}.} The main motivation for adopting the epistemic conception of states is that it allows a reading of the quantum mechanical formalism that, as will be discussed in Section 2, avoids both the quantum mechanical measurement problem and the notorious difficulties connected to quantum ``non-locality'', i.\,e. the apparent tension between quantum mechanics and relativity theory.

Accounts based on the epistemic conception of states belong to two different types: Those of the first type are hidden variable models where the state $\psi$ (or density matrix $\rho$) expresses incomplete information about the configuration of hidden variables that obtains, also called the ``ontic'' state of the system. The defining characteristic of these models, which Harrigan and Spekkens call ``$\psi$-epistemic''\footnote{See \cite{Harrigan}.}, is that an ontic state is compatible with several quantum states $\psi$. Harrigan and Spekkens recommend the search for hidden variable models of this type as a promising alternative to the development of the more traditional ``$\psi$-ontic'' approaches.\footnote{The claim that $\psi$-epistemic models merit close attention is substantiated by Spekkens in \cite{Spekkens}
.} Although this is an intriguing proposal, hidden variable accounts based on the epistemic conception of states will not be the topic of this paper.

The accounts that will be discussed in this text belong to a second type of approach. Accounts of this type are based on the hope that if one adopts the epistemic conception of states, one can get around the paradoxes of measurement and non-locality without specifying an ``ontic'' state of the system at all.\footnote{See the works of Peierls, Mermin, and the quantum Bayesians Fuchs, Caves and Schack mentioned in footnote \ref{fussnote}.} Clarification of the conceptual nature of states alone, it is hoped, may lead to a perspective on quantum mechanics according to which this theory is fine as it stands and as actually applied by working physicists. According to these accounts, no further interpretive take on quantum mechanics, be it in terms of hidden variables, branching worlds, dynamics of collapse or whatever else, is needed in addition to the epistemic conception of states.\footnote{Fuchs and Peres call this an ``interpretation without interpretation'', see \cite{FuchsPeres}. Having in mind the same type of approach, Marchildon writes that ``[t]he question of the epistemic view [of states] is much the same as the one whether quantum mechanics needs being interpreted.'', see \cite{Marchildon} p.\ 1454.} The perspective on quantum mechanics these accounts try to offer can be called ``therapeutic'': The promise they hold is that we may become ``cured'' from our---supposedly---unfounded worries about foundational issues like the measurement problem simply by adopting a certain perspective on states that is free from what is seen as conceptual confusion.

Since only accounts of this second type will be considered in this paper, any further use of the term ``epistemic conception of states'' presupposes a reading that is not in terms of hidden variables. However, even if one restricts one's attention to accounts of that sort, the statement of the epistemic conception of states still allows for various different readings. The aim of the present paper is to find out which account (of the second type) that is based on the epistemic conception of states is the best. Hence, I shall not be trying to defend the epistemic conception of states against the more general objections that have been levelled against it so far.\footnote{For critical voices see \cite{Hagar}, \cite{Marchildon}, \cite{Ferrero}, \cite{HagarHemmo}, \cite{Timpson}.} Nor will I compare its virtues and weaknesses to those of any rival interpretations of quantum mechanics, and I don't intend to suggest in any of the considerations which follow that it has to be preferred over them. The more modest aim of this paper is to determine the version of the epistemic conception of states that \textit{should be} taken as a basis when one compares it to the other philosophical takes on quantum mechanics.

The paper is organised as follows: Section 2 recapitulates some considerations that suggest adopting an epistemic account of states in order to dissolve the measurement problem and the apparent tension between quantum mechanics and relativity theory. In Section 3, I investigate \textit{in which sense} quantum states might reasonably be said to reflect the assigning agents' epistemic relations to the systems. In this context, quantum Bayesianism is discussed, the account developed by C. A. Fuchs, C. M. Caves and R. Schack that analyses quantum probabilities as subjective degrees of belief.\footnote{See the references given in footnote \ref{fussnote}.} It is argued that this position, despite its impressive achievements such as making sense of the practice of quantum state tomography without relying on the notion of an ``unknown quantum state'', is ultimately unsatisfying unless supplemented by an account that rehabilitates the notion of a state assignment being performed \textit{correctly}. In Sections 4 and 5 a novel approach to the epistemic conception of states is proposed the core idea of which is to understand the rules governing state assignment as ``constitutive rules'' in the sense of John Searle. Section 6 discusses to what degree instrumentalism is embraced in that account and to what degree realism remains available.

\section{Motivating the epistemic conception of states}

The most severe difficulty in the foundations of quantum mechanics, at least according to the majority of researchers in the field, is the quantum measurement problem, the fact that if quantum states are seen as descriptions of quantum systems whose time-evolution always follows the Schr\"odinger equation, measurements rarely have outcomes.\footnote{At least if one assumes, as usual, the so-called eigenstate-eigenvalue link which says that for a system in a state $\psi$ an observable $A$ has a definite value $a$ if and only if $\psi$ is an eigenstate of (the operator corresponding to) $A$ with eigenvalue $a$.} If one wants to solve this problem in the spirit of the ``therapeutic'' perspective mentioned before, a natural first step to make is to look at how physicists actually treat the time-dependence of states assigned to systems being measured.\footnote{The idea that a proper treatment of philosophical problems should begin with paying careful attention to the practices (``forms of life'') in which these problems seem to arise is at the heart of the philosophical method of the later Wittgenstein. For remarks on his ``therapeutic'' approach to philosophical questions see, for instance, \cite{PU} \S\S 133, 255.} They resort, of course, to the famous---or, as some would rather say, ``infamous''---von Neumann projection postulate or one of its generalisation, most commonly L\"uders' Rule.\footnote{L\"uders' Rule is given as Eq.\,(\ref{luders}) in Section 5. It can be generalised in different ways, for instance using a POVM-based formulation.}

The projection postulate, however, is not so much liked by everyone. Laura Ruetsche, for example, expresses her qualms about it as follows:
\begin{quote}
Recognizing this [measurement] problem, von Neumann ... responded by invoking the \textit{deus ex machina} of measurement collapse, a sudden, irreversible, discontinuous change of the state of the measured system to an eigenstate of the observable measured. ... Collapse is a Humean miracle, a violation of the law of nature expressed by the Schr\"odinger equation. If collapse and unitary evolution are to coexist in a single, consistent theory, situations subject to unitary evolution must be sharply and unambiguously distinguished from situations subject to collapse. (\cite{Ruetsche} p.\,209)
\end{quote}
Ruetsche brings forward two major points of critique against invoking collapse in order to get around the measurement problem: Her first complaint is that in contrast to the smooth time-evolution governed by the Schr\"odinger equation collapse is sudden and discontinuous, her second complaint is that we are not given any clear criterion for distinguishing between situations where collapse occurs and situations where it does not. She adds that ``despite evocative appeals to such factors as the intrusion of consciousness or the necessarily macroscopic nature of the measuring apparatus, no one has managed to distinguish these situations clearly.''\footnote{See \cite{Ruetsche} p.\,209.} Practising physicists, however, normally seem to know quite well what to do when dealing with systems being measured, even without having a maximally clear-cut criterion for distinguishing between situations where unitary time evolution applies and others where the state must be made to undergo collapse. How do they do it?

The answer is again quite simple: The state of a system, normally evolving continuously according to the Schr\"odinger equation, is subjected to collapse at a time $t_0$ just in case the agent who assigns it to the system takes into account new information about the values of observables with respect to $t_0$. If she has learned, for instance, that the value of an observable $A$ being measured at $t_0$ lies, at that time, between two values $a_0$ and $a_1$, the state assigned for times $t$ immediately after $t_0$ should ascribe probability $1$ to the value of $A$ lying between $a_0$ and $a_1$. If the pre-measurement state does not fulfill this requirement, it has to be adjusted by applying a projection to it. The ambiguity Ruetsche complains of may in practice play a role only as a vagueness about what counts as having obtained information about the values of observables, a vagueness that can be reduced or even completely eliminated by improving either one's method of measurement or one's understanding of the measurement setup. The intuitive motivation for applying collapse, to sum up, is the need of readjusting the state after measurement in order to make it compatible with what one knows of the system as a consequence of the measurement carried out.

Laura Ruetsche, as we have seen, claims that measurement collapse is ``a Humean miracle, a violation of the law of nature expressed by the Schr\"odinger equation''. This view, however, seems compelling and natural only if one thinks of the state and its time-evolution as a description of what actually happens to the system. If, more in line with the original motivation of invoking collapse just sketched, one looks at it as reflecting a sudden change in the state-assigning agent's epistemic situation, not as a sudden change in the system itself, this changes completely. It may therefore be hoped that by adopting a version of the epistemic conception of states, which says that the state reflects the assigning agent's epistemic relation to the system, measurement collapse can be made to look natural.

Of course, this idea is not at all new. That measurement collapse might have to do with a change in what the agent knows (or believes) about the system to which the state is assigned is a thought that occurs to almost any student of quantum mechanics when learning the von Neumann projection rule. Although this reading of collapse is not normally associated to the Copenhagen Interpretation, it has been endorsed by some of the most distinguished proponents of that view.\footnote{It has not, to my knowledge, been defended by Bohr. Faye gives a condensed paraphrase of Bohr's mature view, namely, ``to phrase it in a modern philosophical jargon, that the truth conditions of sentences ascribing a certain kinematic or dynamic value to an atomic object are dependent on the apparatus involved, in such a way that these truth conditions have to include reference to the experimental setup as well as the actual outcome of the experiment.''\cite{Faye} No reference to the epistemic situation of an agent assigning a state is made. Nevertheless, implicit consent to the epistemic conception of states has been attributed to Bohr, see \cite{Merminnew} p. 521.} Heisenberg, for instance, articulates it as follows:
\begin{quote}
Since through the observation our knowledge of the system has changed discontinuously, its mathematical representation also has undergone the discontinuous change and we speak of a `quantum jump'. (\cite{Heisenberg} p.\,28)
\end{quote}
Another important proponent both of the Copenhagen Interpretation and of the idea that the state reflects the assigning agent's epistemic relation to the system is Rudolf Peierls. Peierls writes that the state ``represents our \textsl{knowledge} of the system we are trying to describe'' and he adds that the states assigned by different observers ``may differ as the nature and amount of knowledge may differ''.\footnote{See \cite{Peierls} p.\,19. Peierls, when he talks of the system ``we are trying to describe'', uses the word ``describe'' in a broader sense than I do. According to how ``describe'' is employed in this paper, if one claims about states that they ``describe'' (the properties of) quantum systems, this implies that for each quantum system there is at most one state by which it is correctly described. According to this more narrow usage of ``describe'', Peierls would have had to speak of ``the system we are assigning a state to'', not ``the system we are trying to describe''.}

The importance of this last point can hardly be overstated: If we take seriously the idea that the state reflects the assigning agent's epistemic relation to the system, this has the consequence that it will necessarily differ from agent to agent, depending on the different agents' differing epistemic conditions---and legitimately so. Thus, according to the epistemic conception of states there can be no such thing as an agent-independent ``true'' state of the system---at state it ``is in''---, for if such a state existed, one would need to assign this state in order to assign correctly and assigning any other state would be wrong. Peierls is therefore right to conclude that if indeed the state assignments of different agents having different knowledge of the same system are supposed to reflect the various agents' epistemic relations to the system, it must be allowed that their states may be different. An important aspect of this insight is that it gives us a clear criterion of what we can count as an epistemic account of states: Accounts that acknowledge existence of the \textit{true} state of a quantum system---a state it \textit{is in}---are not varieties of the epistemic conception of states.

Further support for the epistemic conception of states comes from considerations on systems that exhibit what is usually called ``quantum non-locality''. Consider, for instance, a system that consists of two particles that have been prepared in such a way that agents knowing about the preparation procedure will assign an entangled state to the combined system, e.\,g. the state $\frac{1}{\sqrt 2}\left( |+\rangle_A|-\rangle_B\,-\,|-\rangle_A|+\rangle_B \right)$ for the spin degrees of freedom of the combined system. As usual in discussions about non-locality, one assumes that the two systems have been brought far apart and considers an agent Alice, located at the first system and performing a measurement of spin in a certain direction. Having registered the result, Alice will afterwards assign two distinct and no longer entangled states to the two systems, which in general will depend both on the choice of observable measured and on the measured result. Consequently, due to the measurement performed by Alice at the first system the state she assigns to the second system will in part depend on her choice of direction of spin measured at the first system, although the two systems are assumed to be located as far apart as one might wish. On a non-epistemic reading of quantum states---an ``ontic'' reading---this is mysterious because then the state of the second system seems to have undergone an instantaneous change as a consequence of the measurement carried out arbitrarily far away at the first system.\footnote{The same considerations in favour of an epistemic, non-ontic reading of states can be found in more detail in \cite{Fuchs}, Section 3. Fuchs traces the argument to the one developed by Einstein to support a view of states as \textit{incomplete descriptions} of quantum system. In view of the famous no-go results by Bell, Gleason, Kochen and Specker (which were only obtained after the time of Einstein), however, Fuchs sees this route as essentially blocked (see, however, the work of Spekkens mentioned in Section 1). He concludes that Einstein's line of thought provides support for epistemic accounts of states that are not in terms of hidden variables.}

The problem becomes most dramatic in a situation where measurements are performed at both systems by different agents in such a way that the distance between the events (or processes) of the measurement interactions is spacelike, perhaps even in such a way that each measurement is carried out first in its own inertial rest frame.\footnote{For a detailed description of this kind of setup see \cite{Zbinden}. Experimental results reported there confirm the predictions of standard QM.} Here it becomes practically impossible to regard collapse as a real, physical process because there is no non-arbitrary answer to the question of which measurement occurs first and triggers the abrupt change of state of the other.

If one adopts the epistemic conception of quantum states, however, no paradoxical conclusions arise and the sudden change of the state assigned to the second system by Alice appears very natural: During the preparation procedure the two systems have been brought into contact, and because Alice is informed about this, it is unsurprising that the result of her measurement of the system close to herself may affect her epistemic relation to the other. If we interpret the state not as a description of the system but as reflecting her epistemic condition, we are not forced to assume that her measurement has a physical effect on the second system. No change of physical quantities at superluminal velocity needs to be assumed. The predictions of quantum mechanics, based on entangled states, may still be unexpected and surprising, but no conflict with the principles of relativity theory arises.

Before closing this section, it should be pointed out that there is a price to be paid for this elegant dissolution of the tension between quantum non-locality and relativity theory, namely that one has to remain silent about under which conditions which observables are having definite values. Since the notion of an ontic state of the system is not part of the accounts we are concerned with, one cannot, neither by appeal to the eigenstate-eigenvalue link nor to some other, maybe more sophisticated rule, define criteria of under which conditions an observable has a determinate value in terms of the state of the system. As long as one does not introduce an additional ontic state of the system---as in the hidden variable accounts mentioned in Section 1---there is probably no way of doing so at all. Epistemic accounts of the quantum state which are not in terms of hidden variables must therefore be based on a more pragmatic point of view according to which quantum mechanics makes statements (or predictions) about the values of observables only insofar as these values might indeed be determined, registered or otherwise encountered by agents. The instrumentalist spirit behind this presupposition may seem questionable, but given how elegantly the epistemic conception of states dissolves the paradoxes of measurement and non-locality, there are good reasons for taking the view very seriously nonetheless. Furthermore, as I shall argue in the last section of this paper, the version of the epistemic conception of states developed in Sections 4 and 5 of this paper is compatible with a substantial amount of realism.

\section{Knowledge of probabilities vs. probabilities as degrees of belief}

A quantum state, via the Born rule, assigns probabilities to the different possible values of observables of a quantum system. It is only in virtue of these probabilities---or, alternatively, expectation values---that quantum mechanics is empirically testable. Since assigning a state to a system means assigning probabilities to the values of observables, it may seem natural to read Peierls' claim that the state ``represents our knowledge of the system'' as a shorthand for saying that the state represents our knowledge of these probabilities. Peierls' position, interpreted along these lines, is sometimes even straightforwardly identified with the epistemic conception of states, for example by Marchildon, who claims that ``[i]n the epistemic view [of states], the state vector (or wave function or density matrix) does not represent the objective state of a microscopic system [...], but rather our knowledge of the probabilities of outcomes of future measurements.''\footnote{See \cite{Marchildon} p.\ 1454.} However, as has been suggested by Fuchs\footnote{See \cite{Fuchs}, Footnote 9 and Section 7 in particular. The argument given in the following section can also be found in \cite{Timpson}, Section 2.3.}, the notion of knowledge of quantum probabilities is incompatible with the epistemic conception of states, so to define it in terms of this notion is highly problematic.

To see why there can be no knowledge of quantum probabilities in an epistemic account of states, recall that it is a crucial feature of the notion of knowledge that it is \textit{factive}. This means that, according to what the term ``knowledge'' means, it is impossible to know that $q$ unless the sentence ``$q$'' is indeed true. Due to this aspect of ``knowledge'' the view described by Marchildon---that the state reflects our knowledge of probabilities---is incompatible with what we determined to be an essential ingredient of the epistemic conception of states, the assumption, namely, that different agents may legitimately assign different states to one and the same system. For assume that probabilities are indeed the objects of our knowledge so that an agent might know the probability $p$ of a certain measurement outcome $E$ to occur. In this case an immediate consequence of the fact that ``knowledge'' is factive would be that $p$ is the one and only correct, the \textit{true} probability for $E$ to occur. Since this line of thought applies for any possible measurement outcome $E$ some probability $p$ is ascribed to by the state assigned to the system, we are bound to conclude that the probabilities obtained from this state are the (only) true ones so that any other assignment of probabilities would simply be wrong. But this is incompatible with the fundamental assumption of the epistemic conception of states that the states assigned by different observers, as Peierls writes, ``may differ as the nature and amount of knowledge may differ''. Consequently, saying that quantum states represent our knowledge of quantum probabilities is not an option for spelling out the epistemic conception of states.

Nevertheless, adherents of the epistemic conception of states may assert that quantum states reflects the assigning agents' subjective \textit{degrees of belief} about possible outcomes. The resulting version of the epistemic conception of states has been worked out by Fuchs, Caves and Schack and is known as \textit{quantum Bayesianism}. According to this view, quantum states reflect not our knowledge, but rather our beliefs about what the results of ``our interventions into nature''\footnote{This is how Fuchs describes it, see \cite{Fuchs} p.\,7.} might be. Quantum probabilities, in this view, are not the objects of our belief, but they indicate how strongly we believe that the measurement outcomes in question might occur.

Quantum Bayesianism, by interpreting quantum probabilities as subjective degrees of belief, does not run into any problems like the one described that arises for the view that quantum states represent our knowledge of probabilities: Degrees of belief may differ from agent to agent, without one of them necessarily being in error or making a mistake. Applied to the quantum mechanical case, this means that quantum Bayesianism can nicely allow for the possibility that, as Peierls writes, the states assigned by different agents ``may differ as the nature and amount of knowledge may differ''. As far as I see, the position is therefore consistent as an attempt to provide a version of the epistemic conception of states. Timpson goes as far as concluding  that if one wants to spell out the meaning of `state' ``in terms of some cognitive state... quantum Bayesianism, where the cognitive state called on is belief, not knowledge, is the only consistent way to do that''.\footnote{See \cite{Timpson} p.\,593.}

However, quantum Bayesianism goes extremely far in characterising elements of the quantum mechanical formalism as subjective in order to be consistent as an epistemic account of states. How radical the view really is becomes strikingly clear from the fact that, for any given measurement device, quantum Bayesianism denies the existence of a determinate answer to the question of \textsl{which} observable is measured in that setup. As explained by Fuchs:
\begin{quote}
Take, as an example, a device that supposedly performs a standard von Neumann measurement $\{\Pi_d\}$, the measurement of which is accompanied by the standard collapse postulate. Then when a click $d$ is found, the posterior quantum state will be $\rho_d=\Pi_d$ regardless of the initial state $\rho$. If this state-change rule is an objective feature of the device or its interaction with the system---i.\,e., it has nothing to do with the observer's subjective judgement---then the final state must be an objective feature of the quantum system. (\cite{Fuchs} p.\,39)
\end{quote}
Fuchs' reasoning can be summed up as follows: If it is an objective feature of the device that it measures an observable having a set $\{\Pi_d\}$ of one-dimensional projection operators as its spectral decomposition, then, as soon as a ``click $d$'' has been registered, the state that must be assigned after measurement is $\Pi_d$ independently of the state $\rho$ which has been assigned before. No freedom of state-assignment remains in this case and we seem to have ended up with an agent-independent true post-measurement state $\Pi_d$ the existence of which is incompatible with the epistemic conception of states we wanted to spell out in detail. As a consequence, the conclusion drawn by Fuchs---that there can be no objective fact of the matter concerning which observable is measured by which measurement device if one assumes the epistemic conception of states---seems hard to avoid.

Although this conclusion may---wrongly, as I shall argue---seem unavoidable, it is nevertheless extremely hard to accept, and some may even regard it as a \textit{reductio} of the whole project of giving an epistemic account of states. Agreement on which observable is measured in which experimental setup is pervasive among competent experimentalists, and it seems difficult to imagine how quantum mechanics could be empirically successful if it were not. Furthermore, if there were no fact of the matter concerning which observable is measured in which context, there could also be no fact of the matter concerning \textit{which} observable some numerical value obtained in an experiment is a value \textit{of}. Hence, it would quite generally be impossible to know the values of any observables because one could never know to which observable some given value really belongs. Since it is hard to deny that we often do have knowledge of the values of at least some observables, this consequence of Fuchs' reasoning is extremely problematic. Even if one adopts the quite radical view that the observables of microscopic systems (whatever one counts as such) never have determinate values, one can hardly make the same claim for those (normally macroscopic) systems to which we have more direct access and which are also treated quantum mechanically by means of the many-particle methods of quantum statistical mechanics (e.\,g. when computing heat capacities, magnetic susceptibilities, and the like). Here it is usually assumed that one has at least approximate knowledge of the volume, particle number, temperature, maybe pressure of the (macro-) system, which gives constraints on the values of observables of the individual (micro-) particles.

However, the conclusion that we cannot obtain any knowledge of the values of observables seems contrived even with respect to microscopic systems that are not part of a many-particle macrosystem. If we consider, for instance, a Stern-Gerlach setup that measures the $x$-component of electron spin, it seems implausible to hold that even with respect to the moment of measurement the experimentalist is in principle unable to know the value of (the $x$-component of) spin, which would be the case if there really were no fact of the matter as to which observable she has measured. When we look at how physicists actually talk and behave, we have, as it seems, strong evidence that obtaining knowledge of the values of observables is possible. I am not thereby saying that quantum Bayesianism owes us an ``ontological'' account of under which conditions which observables do have determinate values, for this, as argued before, is something that can hardly be expected from an epistemic account of states that is not in terms of hidden variables. Something much more modest is asked for, namely that the mere possibility of having knowledge of the values of observables should not be ruled out as a matter of principle.

A final drawback of the view that the question of which observable is measured in which setup has no determinate answer is that it undermines the notion of a state assignment being performed \textit{correctly}, which is also essential for quantum mechanical practice. In the case of systems having been prepared by a (so-called) \textit{state preparation device}, for example, any state assignment that deviates from a highly specific one can reasonably be counted as wrong. Since state preparation can be regarded as a form of measurement, allowing the question of which observable is measured in which setup to have a determinate answer is the same as acknowledging the notion of a correct (or incorrect) state assignment.\footnote{As will be the focus of Section 4 and 5, this is not the same as acknowledging the notion of a state the system \textit{is in}.} Quantum Bayesianism, arguably, should find a way of making sense of that notion.

There is a strategy to which defenders of quantum Bayesianism might try to resort in order to defend their account against the objection just presented, namely by trying to offer a story that explains why physicists may \textit{in practice} behave as if the observable measured were determinate although, in reality, this is not so. They might try model such a story on the account they have developed to explain why physicists' talk of ``unknown quantum states'' (as one finds it, for example, in quantum state tomography) successfully serves the purposes of communication, although, from the quantum Bayesian point of view, the notion of an ``unknown quantum states'' is illegitimate. The argument given by Caves, Fuchs and Schack invokes a variant of de Finetti's classical representation theorem on sequences of events that are subjectively judged to be ``exchangeable''.\footnote{See \cite{Caves2002b}.} On the basis of this theorem, it becomes understandable why different agents who register the same measured data will come to agreement in their state assignments even if the states they started out with are very different. The sole presupposition for this to happen is that they (subjectively) judge the states of the sequence of measured systems to be \textsl{exchangeable}, roughly meaning that for the assigning agents both the individual positions of the systems in the sequence of trials and the number of measurement trials carried out play no role.\footnote{For the precise definition of an exchangeable sequence of states see \cite{Caveserratum}, an erratum note to \cite{Caves2002b}.} As demonstrated by Caves, Fuchs, and Schack, this assumption is sufficient to enforce that ``the updated probability $P(\rho|D_K)$ becomes highly peaked on a particular state $\rho_{D_K}$ dictated by the measurement results, regardless of the prior probability $P(\rho)$, as long as $P(\rho)$ is nonzero in a neighborhood of $\rho_{D_K}$.''\footnote{See \cite{Caves2002b} p.\,4541.} Hence, this theorem can be used to explain why the states assigned by different agents starting from different priors will become practically indistinguishable after a sufficiently large number of experiments witnessed without there being any such thing as the ``unknown state'' any of the systems really is in.\footnote{For an analogous result concerning quantum \textit{process} tomography see \cite{FuchsSchackScudo}.}

This argument impressively shows how talk about ``unknown quantum states'' can be given an interpretation according to which it no longer includes a commitment to states as descriptions of quantum systems. However, there is no analogous way to interpret (and thereby justify) talk about ``the observable $A$ measured by a certain device $D$'' avoiding commitment to the determinateness of observables measured. There is at least one highly important difference between the two cases, namely that while we update our state assignment after having registered a measurement result, there is no prescription for adjusting our beliefs about the observable measured. So, there can be no derivation of a de Finetti-type reconstruction of talk involving ``the observable measured'' along similar lines as the one involving talk about ``unknown quantum states''. It should also be noted that agreement on which observable has been measured has to be presupposed in the quantum Bayesian take on ``unknown quantum states'': Measurement results could never ``dictate'' a certain state that must be assigned after a certain number of measurement trials unless we assume that in each case the observable measured is an objective matter. The assumption that different observers agree on which observable is measured in which case crucially enters the reasoning offered by Caves, Fuchs and Schack, but we are given no explanation of how this agreement might come about.

What is needed to improve upon quantum Bayesianism as an epistemic account of states is a justification of talk about measurement results ``dictating'' state assignments without thereby reintroducing the notion of a state the system is in. In order to see whether such an account can be given, we must reexamine the claim made by Fuchs that in an epistemic account of states there can be no determinate answer to the question of which observable is measured by which setup.

\section{Objectivity of observables measured in an epistemic account of states}
According to Fuchs, there can be no determinate answer to the question of which observable is measured in which experimental setup, because if the observable measured were an objective feature of the device, the measured result would impose objective constraints on the state that has to be assigned to the system after measurement. Fuchs sees a conflict between this conclusion and the basic assumption of the epistemic conception of states that there is no agent-independent \textit{true} state of the system. However, as I shall try to show now, this line of reasoning is not cogent. The idea of combining an epistemic account of states with the view that to a given setup there corresponds a determinate observable that is measured in that setup is perfectly coherent.

To see why this is the case, let us assume, in accordance with the epistemic conception of states, that the state one has to assign to a system should somehow depend on the information one has about the system. To this we now add the assumption that the measured observable is an objective feature of the setup and that agents registering measurement data may obtain information about the value of the observable that is measured.

Now, does it really follow from these assumptions that there is a certain state the system is in after measurement---the \textit{true} state of the system, as one might call it. Clearly no: All that follows is that agents having registered a result must update the states they assign in a way that depends on the observable measured together with the registered result. As regards the case described by Fuchs, those having registered the ``click $d$'', according to the assumptions made, are obliged to assign $\Pi_d$ as their post-measurement state in order to perform their state assignments correctly. This, however, does not mean that $\Pi_d$ must be regarded as the \textit{true} or \textit{real} post-measurement state of the system, the one it really \textit{is in} after measurement. To reach this further conclusion, something more would have to be shown, namely that assigning a state that is different from $\Pi_d$ would be wrong irrespectively of what one may know of the system, i.\,e. wrong not only for those who know about the ``click $d$'', but also for others who don't.

So let us consider more closely the situation of agents who are assigning states to the system---normally different from $\Pi_d$---without having had a chance to register the ``click $d$''. This may even be a matter of physical impossibility, due to the fact, namely, that the measurement process (or event) resulting in $d$ lies outside their present backward light cone. Now, if we take seriously the idea that the states these agents assign reflect their epistemic situations with respect to the system, does it make sense to claim that, nevertheless, they are wrong to assign the states they assign, which are different from $\Pi_d$? Clearly not: Their epistemic relations to the system, by hypothesis, are such that they have no reason at all for assigning $\Pi_d$. If their state assignments should indeed be adequate to their epistemic relations to the system, assigning $\Pi_d$ would not only not be mandatory for them, it would even be \textit{wrong}, for it would not conform to the knowledge they have of the values of observables of the system.

Assuming that the observable measured is an objective feature of the device, we see, does not lead to the conclusion that there is a uniquely distinguished post-measurement state that must be assigned by anyone who intends to assign correctly. Consequently, determinateness of the observable measured does not imply that there is an agent-independent \textit{true} state the system is in after measurement. We can therefore conclude that the argument given by Fuchs falls short of establishing that there can be no fact of the matter as to which observable is measured in which case if one assumes the epistemic conception of states.

It might be objected against this line of thought that it accords too much importance to the states assigned by those who are simply not well-informed about recent measurement results concerning a system. There are cases when measurement outcomes narrow down possible state assignments to a unique \textit{pure} state, and this state, so the objection might go, clearly has a special status. Assignment of it is based on the best possible knowledge of the values of observables to be had and refusing to call it the state the system really is in may therefore seem artificial. Why should one accord any significance to the state assignments of those whose epistemic situation with respect to the system is simply worse?

To answer this challenge, it should first be emphasised that the epistemic conception of states has no problem to admit that knowledge of the values of observables can be better or worse (in the sense of, say, more or less detailed, more or less up to date etc.) and that a state assignment based on excellent knowledge of the values of observables is likely to lead to the best predictive results. In that sense, when there are agents having knowledge of the values of observables which cannot be further improved (at least not without losing part of this knowledge by measuring the system again) and narrows down possible states to assign to a unique pure state, there is nothing wrong with regarding this state as enjoying a special and privileged status. There is no need, however, to conclude from this that this pure state corresponds to a physical property of the system, one that it has independently of someone being there who happens to assign it. We are free to regard the expression ``state assigned by those whose knowledge about the values of observables cannot be further improved'', in case it refers to some state at all, as having a referent only because an agent\footnote{Here one might think not only of a human agent but also of an artificially constructed device, provided that it is capable of registering and storing information and can meaningfully be described as assigning a quantum state to a quantum system.} happens to be there who actually has such excellent knowledge. On some occasions (on ``state occasions'', one might say) there can be agents having knowledge of the values of observables which narrows down possible states to assign to a uniquely determined pure state. The assumption, however, that even when there are no agents having knowledge of that sort, there exists some state which \textsl{would have to be assigned} by anyone assigning a state to the system need not be made and does not go well with the epistemic conception of states. Consequently, by conceding that from time to time agents may have knowledge of the values of observables which cannot be further improved one is not committed to the view that for any system there exists some state it is in. Trying to spell out the epistemic conception of states, we can therefore allow that the question of which observable is measured in which setup may have a determinate answer, and we do not have to hold that knowledge of the values of observables is impossible in principle, as follows from the reasoning given by Fuchs.

Once we accept the view that the question of which observable is measured in which setup has a determinate answer, we are in a position, unlike the proponents of quantum Bayesianism, to make sense of the notion of a state assignment being performed correctly. From the perspective of the standard conception of states where states are seen as descriptions of quantum systems this may seem puzzling: According to this conception, a state assignment is correct if and only if it is an assignment of the state the system really is in. However, as I shall show in the following section, saying that a state assignment has been performed correctly remains coherent even if one rejects the notion of a state the system is in.

\section{Constitutive rules}
In order to preserve the notion of a state assignment being performed correctly without relying on the notion of a state a quantum system is in, we have to appeal to the rules employed in the assignment of quantum states, arguing that assigning correctly means to assign in accordance with them. Assuming the epistemic conception of states to be valid, we should think of these rules as determining the state an agent has to assign to the system as a function of her knowledge of the values of its observables. Let me briefly discuss the most basic examples of such rules by going through the different types of situations in which they apply.

The simplest type of situation we have to consider is when no new knowledge of the values of observables is obtained. In these cases, the state of the system must be evolved in time following unitary time-evolution as determined from the time-dependent Schr\"odinger equation. What is normally seen as a \textit{fact} about quantum states---that their time-evolution follows the Schr\"odinger equation---takes the form of a rule of state assignment, the rule, namely, that an agent should apply unitary time-evolution for all times $t$ with respect to which she has no new incoming data concerning the values of observables of the system.

The prescription that has to be used in case our agent \textit{does} acquire new knowledge of the values of observables, namely that, to be specific, at a certain time $t_0$ the value of an observable $A$ lies within a certain range $\Delta$, is L\"uders' Rule. L\"uders' Rule can be motivated as the analogue of Bayes' Rule for probability conditionalisation in the light of new evidence in a non-commutative setting.\footnote{See \cite{Bubold} and \cite{Bub}. For a proposal of how to interpret L\"uders' Rule as the quantum analogue of Bayes' Rule in the context of the quantum Bayesian framework, see \cite{Fuchs} Section 6. More complicated versions of the measurement update rule are required when dealing with unsharp measurements or generalised measurements using the formalism of POVMs.} If we denote by $\rho$ the state assigned to the system immediately before $t_0$ and by $\Pi_{\Delta}$ the projection onto the linear span of eigenvectors of $A$ with eigenvalues lying within $\Delta$, the change of state according to L\"uders' Rule is given by

\begin{equation}\label{luders}
\rho\longrightarrow\rho_\Delta=\frac{\Pi_{\Delta}\rho\Pi_{\Delta}}{\rm{Tr}\left(\Pi_{\Delta}\rho\Pi_{\Delta}\right)}\,.
\end{equation}

However, both unitary time-evolution in accordance with the Schr\"odinger equation and collapse in accordance with L\"uders' Rule are rules of state change. They can be applied to the state of the system only if a state has already been assigned in the first place. We might, however, also be interested in the standard of correctness for the assignment of states to systems where no state has been assigned before. In some cases this problem can be solved by appeal to L\"uders' Rule alone, namely when one's knowledge of the values of observables uniquely fixes the post-measurement state so that the pre-measurement state---if one had been assigned---would not have any influence on the post-measurement state.

In the generic case, however, this is not sufficient to determine the state one has to assign. In quantum statistical mechanics, for example, one is usually dealing with systems for which one has knowledge of only a very limited number of quantities, typically called ``macroscopic variables'' such as temperature, pressure, magnetisation, etc. Here one expects that the state to be assigned should conform to the criterion that it maximises entropy subject to certain constraints that are determined from what one knows of the values of these variables.\footnote{Although, as mentioned before, L\"uders' Rule can in some special cases replace entropy maximisation, it should be noted that in general Bayesian conditionalisation (i.\,e., in the quantum context, L\"uders' Rule) and entropy maximisation serve different purposes and should not be seen as competing principles. For an instructive assessment of their differing roles see \cite{Jaynesnew}.} Entropy maximisation, as emphasised by Jaynes, can be motivated on grounds that it leads to ``the only unbiased assignment [of probabilities] we can make; to use any other would amount to arbitrary assumption [\textit{sic}] of information which by hypothesis we do not have.''\footnote{See \cite{Jaynes1} p.\,623.} In quantum mechanics, entropy must be given as a function of the state assigned to the system, and it is widely believed that the von Neumann entropy $S(\rho)=-k_B\rm{Tr}(\rho\log(\rho))$ is the appropriate quantity here. Although there has recently been some debate on whether this view is really correct\footnote{See \cite{Shenker} \cite{Hendersonold}, \cite{hemmoshenker}. For an early but still very useful discussion of whether the von Neumann entropy is the adequate quantity in the quantum mechanical context see \cite{Jaynes2}.}, we need not be concerned with this question here. It is sufficient for us to assume that an entropy function exists that has to be maximised when a state is assigned to a system to which no state has been assigned before.

Unitary time-evolution, L\"uders' Rule and entropy maximisation have been discussed here although they are surely familiar because the \textit{status} ascribed to them in the epistemic account of states proposed here is quite unusual. It is, in particular, very different from the status these principles have in the standard conception of states as descriptions of quantum systems. The nature of this difference can be clarified by bringing into play some terminology invented and introduced by John Searle in the context of his theory of speech acts.\footnote{See \cite{Searle}.} Searle distinguishes between two different sorts of rules, and this distinction is very useful for clarifying the role of the rules of state assignment in the epistemic account of states proposed here.

The distinction between the two kinds of rules is introduced by Searle as follows:
\begin{quote}
I want to clarify a distinction between two different sorts of rules, which I shall call \textit{regulative} and \textit{constitutive} rules. I am fairly confident about the distinction, but do not find it easy to clarify. As a start, we might say that regulative rules regulate antecedently or independently existing forms of behavior; for example, many rules of etiquette regulate inter-personal relationships which exist independently of the rules. But constitutive rules do not merely regulate, they create or define new forms of behavior. The rules of football or chess, for example, do not merely regulate playing football or chess, but as it were they create the very possibility of playing such games. The activities of playing football or chess are constituted by acting in accordance with (at least a large subset of) the appropriate rules. Regulative rules regulate pre-existing activity, an activity whose existence is logically independent of the rules. Constitutive rules constitute (and also regulate) an activity the existence of which is logically dependent on the rules. (\cite{Searle} p.\,33 f.)
\end{quote}
According to the standard view of quantum states as states quantum systems ``are in'', the rules of state assignment are regulative rules. To see this, assume that a system is indeed correctly described by an agent-independent true state. The existence of such a state, according to this perspective, is independent of whether there are any agents who might actually happen to assign it. Therefore, what agents are aiming at when assigning a state---namely, to assign the state the system really is in---can be specified without mentioning the rules one follows in order to achieve this goal. Consequently, according to the standard view of states as descriptions of quantum system, the notion of a state assignment being performed correctly is ``logically independent of the rules''\footnote{Phrases within quotation marks in this and the following paragraphs are all taken from the passage from \cite{Searle} just cited.} according to which it is done, namely those discussed before in this section. If states are seen as describing the properties of quantum objects, the rules of state assignment play the role of an instrument or a guide that is used to arrive at the true state of the system (or some reasonable approximation to it). State assignment, from this point of view, is a ``form of behavior'' that makes sense and therefore exists ``antecedently [to] or independently'' of the rules according to which it is done. It is \textit{regulated} by these rules without being \textit{constituted} by them, in the sense these expressions have in the writings of Searle.

In the version of the epistemic conception of states developed here, however, the status of the rules of state assignment is very different: Their role cannot be that of a guide or instrument to assign a state that is hoped to be (a good approximation to) the \textit{true} state of the system, for the notion of such a state is rejected. Rather, assigning a state in accordance with the rules of state assignment is what it \textit{means} to perform a state assignment correctly, so the notion of a state assignment being performed correctly is itself \textit{defined} in terms of these rules. Consequently, the notion of a state assignment being performed correctly is ``logically dependent on the rules'' according to which it is done. We therefore have to conceive of the rules of state assignment as constitutive rules if, in an epistemic account of states, we want to save the notion of a state assignment being performed correctly without being forced to accept the notion of a state a quantum system is in.

\section{The charge of instrumentalism}

It can hardly be denied that any epistemic account of states that is not in terms of hidden variables, including the one proposed here, does not live up to the expectations of those looking for what may be called a ``robustly realist'' view of quantum mechanics. To some extent this is probably a necessary price to pay if one wants to understand quantum mechanics in line with the more ``therapeutic'' attitude described in Section 1. There is a tension between accounts that propose a non-descriptive reading of states and philosophical realism because the ambition to interpret physical theories as describing the world (as it \textit{really} is) belongs to the core of realist thought. Thus, adopting an epistemic account of states means that from the very beginning one has to make some concessions to the anti-realist or, more specifically, instrumentalist view that physical theories are essentially (nothing more than) computational tools to predict the outcomes of experiments. An understanding of quantum mechanics that accepts this doctrine in its pure and most radical form is (trivially) possible, but at the same time completely uninteresting. The version of the epistemic conception of states proposed here, however, is arguably not committed to such an extreme instrumentalism. Before closing the paper, I will briefly discuss how much of instrumentalism is unavoidably present in the account proposed here and how much of realism remains compatible with it.

A highly characteristic feature of the perspective on quantum mechanics taken by the epistemic conception of states is that according to this perspective, as Timpson formulates it, ``the theory does not proffer a view from nowhere''\footnote{See \cite{Timpson} p.\,591.}, where the phrase ``view from nowhere'' has been drawn from the title of the famous book \cite{Nagel} by Thomas Nagel. According to the epistemic conception of states, it is wrong to think of quantum systems as being ``in'' quantum states. Any state must be thought of as being assigned to the system by an agent having certain knowledge of the values of observables of the system. Since the formalism of quantum mechanics makes empirically testable claims on the basis of quantum states, this agent-dependence carries over to the formalism as a whole: According to the epistemic conception of states, the whole formalism has empirical meaning (or content) only insofar as it may in fact be applied by an agent who can be characterised by his having a certain epistemic condition. This is very different from the picture offered in the more pronouncedly realist interpretations of quantum mechanics such as, say, the Everett interpretation, which assumes that the universe (or ``multiverse'', according to some proponents) is described by a wave function, which is accessible only from something like a God's eye point of view.

Denying that quantum mechanics delivers a ``view from nowhere'' at the world, the epistemic account of states proposed here is surely less ``realist'' than those might have hoped who are looking for a fully-fledged realist view of the theory. However, a strategy for remaining as realist as possible about quantum mechanics while keeping an epistemic account of states is described by Timpson in his study of quantum Bayesianism:
\begin{quote}
Quantum mechanics may not be a descriptive theory, we may grant, but it is a  significant feature that we have been driven to a theory with just this characteristic (and unusual) form in our attempts to deal with and systematize the world. The structure of that theory is not arbitrary; it has been forced on us. Thus by studying in close detail the structure and internal functioning of this (largely) non-descriptive theory we have been driven to, and by comparing and contrasting with other theoretical structures, we may ultimately be able to gain \textit{indirect} insight into the fundamental features of the world that were eluding us on any direct approach; [...]\\
Thus the essence of the quantum Bayesian position is to retain a realist view of physics and of the world whilst maintaining that our fundamental theory---quantum mechanics---should not itself receive a realist reading; while no simple-minded realist alternative is to be had either. (\cite{Timpson} p.\,582 f.)
\end{quote}
The main idea of Timpson's proposal of how to combine ``realism about physics'' with a view of quantum mechanics as non-descriptive is to regard the ``structure and internal functioning'' of the theory as somehow corresponding to the facts of nature. Another way of conveying the same idea would be to say that these aspects of the theory have been \textit{discovered} by physicists and are not merely free creations or inventions of the human mind. The idea can be made a little more specific by invoking a proposal by Jeffrey Bub, who suggests to interpret the fact that the lattice of projection operators on the Hilbert spaces employed in quantum mechanics is non-Boolean as corresponding to ``an objective feature of the world, the fact that events are structured in a non-Boolean way''.\footnote{See \cite{Bub} p.\,252.} The position defended by Bub invites similar charges of being instrumentalist as quantum Bayesianism, for he also interprets quantum mechanical probabilities as (merely) subjective degrees of belief. By means of the cited remark he tries to distance himself from the instrumentalist readings his account of quantum probabilities may receive.

The version of the epistemic conception of states proposed here is in any case more compatible with realist ambitions than quantum Bayesianism because it accepts the notion of a correct (or incorrect) state assignment as meaningful. This allows one to claim that not only the internal structure of the quantum mechanical formalism is non-arbitrary---for example the fact that the lattice of projection operators on a Hilbert space used in quantum mechanics is non-Boolean---, but also the technique of how this formalism is brought in contact with the world. Furthermore, one does not even need two ``separate'' realist strategies for the formalism and the method if one realises that according to the present account the rules of state assignment, which are part of the method, are at the same time part of what constitutes the empirical meaning of the formalism. To see why this is so, recall that the account proposed here regards the rules of state assignment as constitutive rules which determine for linguistic acts involving quantum states whether they qualify as correct or not. Since the empirical meaning of the constituents of the quantum mechanical formalism certainly depend on what counts as correct linguistic usage of them, we arrive at the conclusion, based on the present account, that at least part of the empirical meaning of the quantum mechanical formalism is determined by the rules of state assignment.

Realist accounts of physical theories are expected to apply to these theories not as uninterpreted pieces of pure mathematics but as having a certain empirical meaning. The claim that aspects of the mathematical formalism used to formulate quantum mechanics \textit{somehow} correspond to objective features of the world is not very interesting. It becomes nontrivial only by specifying a certain empirical interpretation for the mathematical formalism in virtue of which it reflects objective features of the world. According to the present account the empirical significance of quantum states is fixed (at least in part) by what counts as correctly assigning a quantum state to a physical system. Consequently, if one wants to combine the epistemic account of states proposed here with elements of realism about quantum mechanics, one should hold that the quantum mechanical formalism reflects objective features of the world in virtue of the empirical meaning of its constituents, which is partly fixed by the rules governing state assignment.

The fact that the present version of the epistemic conception of states is compatible with a non-negligible amount of realism about physics shows that it does not collapse into an uninteresting form of instrumentalism. It is important, however, to keep in mind that the ``therapeutic'' perspective on quantum mechanics which was introduced in the beginning of this paper to motivate the epistemic conception of states is neither realist nor anti-realist in spirit. As explained before, it is based on the ``deflationary'' idea that once the conceptual roles of the elements of the formalism (in particular those of states) have been clarified, quantum mechanics no longer calls for an interpretation (in the sense discussed in Section 1). It is, of course, very difficult and far beyond the scope of the present investigation to assess whether or not this attitude is ultimately adequate and how well it fares in comparison with the stances underlying the more realistically minded interpretations such as Bohmian mechanics, the Everett interpretation or GRWP-theory. If, however, one considers adopting the ``therapeutic'' perspective, I recommend basing one's view on the epistemic conception of states together with an understanding of the rules of state assignment as constitutive rules in Searle's sense.

\end{document}